# Greening Big Data Networks: The Impact of Veracity

Ali M. Al-Salim, Taisir E. H. El-Gorashi, Ahmed Q. Lawey, and Jaafar M. H. Elmirghani

*Abstract*—The continuous increase in big data applications, in number and types, creates new challenges that should be tackled by the green ICT community. Big data is mainly characterized by 4 Vs: volume, variety, velocity, and veracity. Each V poses a number of challenges that have implications on the energy efficiency of the underlying networks carrying the big data. Addressing the veracity of the data is a more serious challenge to data scientists, since they need to distinguish between the meaningful data and the dirty data. In this article, we investigate the impact of big data's veracity on greening IP by developing a Mixed Integer Linear Programming (MILP) model that encapsulates the distinctive features of veracity. In our analyses, the big data network was greened by cleansing the raw big data before processing and then progressively processing the cleansed big data at strategic locations, dubbed processing nodes (PNs). The PNs are built into the network along the path from the sources to the centralized datacenters. At each PN, the cleansed data was processed and smaller volume of useful information was extracted progressively, thereby, reducing the network power consumption. Furthermore, a backup for the cleansed data was stored in an optimally selected Backup Node (BN). We evaluated the network power saving that can be achieved by a green big data network compared to the classical non-progressive approach. We obtained up to 52% network power savings, on average, in the green big data approach compared to the classical approach.

## I. INTRODUCTION

The remarkable evolution of Internet-enabled technologies is driving the world to be inundated by a colossal amount of data generated from various domains, such as bioinformatics, health informatics, social media, text, log files, sensors data, video streaming, purchase transaction records and more. The term big data has been devised to describe the handling of the enormous number of data types generated by numerous data sources.

Although it is currently extremely hard to enumerate the volumes of data generated from a large number of Internet-enabled devices, the situation is going to be more complicated in the near future, as the projected number of Internet-connected devices is anticipated to reach 100 billion devices by 2020 [1]. Based on the International Data Corporation (IDC) report [2], the overall envisaged data volume will reach 40,000 Exabytes in 2020. This exponential increase in the speed of generating data, in the volumes of data and in the variety of big data sources comes in parallel with drops in the percentage of data processed inside datacenters (DC) because of insufficient and inefficient analysis tools [3]. Accordingly, a large amount of the data to be processed is neglected, deleted or delayed. Thus, immense networking power is consumed due to transferring unprocessed data from its sources to DCs, while the only interest is in the small volume of knowledge it carries. Furthermore, extra wastage in storage and bandwidth can result from transferring raw data, which leads to a magnification of the financial costs.

Effort has been directed towards the minimization of communication cost and power consumption when processing and transporting big data. In [4] the authors presented a Mixed Integer Linear Programming Model (MILP) to minimize the overall cost for big data placement, processing, and movement across geo-distributed data centers. The authors in [5] proposed a MapReduce framework to locally process as much data as possible on multiple IoT nodes rather than transmitting the raw data to DCs. The authors in [6] presented a processing system for executing a sequence of MapReduce jobs in Geo-distributed DCs where the processing of jobs is optimized according to time and pecuniary cost. The authors in [7] aimed to satisfy as many big data queries as possible over a number of time slots while keeping the communication cost to a minimum. The authors in [8] proposed in-network processing as a technique to achieve network-awareness to reduce bandwidth usage by custom routing, redundancy elimination, and on-path data reduction. In [9], the authors presented a framework for energy efficient cloud computing services in IP over WDM core networks.

In [10], we presented preliminary results to demonstrate the impact on network power consumption of processing and transferring big data in bypass IP over WDM networks. We considered one big data type from the MapReduce platform that was obtained from the log files of MapReduce clusters from Facebook [11]. In this type, the volume of the output of the reduce process is very small compared to the input of the mapping process. We investigated improving the energy efficiency of big data networks by processing this data type progressively in processing nodes (PNs) of limited processing and storage capacity along the data journey through the IP over WDM core networks to the DCs. The amount of data transported over the core networks was significantly reduced each time the data was processed; therefore, we referred to such a network as an Energy Efficient Tapered Data Network. In [12], we presented three scenarios to further investigate the impact of the progressive processing on green big data networks by serving different input volumes in the network.

The characteristics of big data can be classified into four main Vs: volume (and its effects on the power requirements), velocity (with impacts on the speed of processing), variety (where different applications need various CPU requirements) and veracity (that specifies trustworthiness, data protection and data cleansing and backup constraints).

We developed in [13] and [14] MILP models to investigate the impact of the big data's volume, variety, and velocity on greening big data networks. We presented several scenarios to process different types, volumes, and processing speeds of big data progressively, starting at the edge of the network, moving through the intermediate nodes of the network and finally processing at the central datacentres.

The veracity of big data is a more serious challenge to data scientists since they need to distinguish between the meaningful data and the dirty data [15]. Significant efforts need to be put forward to keep dirty data out of organizations' databases. A good reason to motivate big data scientists to analyze the veracity is that, for example, low-quality data causes the U.S. economy to waste $3.1 trillion each year [15].

Data cleansing [16] deals with detecting and removing errors and duplications from data to improve its quality. When dealing with multiple big data sources, the need for data cleansing becomes significant since the sources may contain dirty data due to overlaps, duplications or contradictory materials. Therefore, it is important to cleanse data so that it is readied for big data analytics. Hence, providing easy access to accurate, consistent and consolidated data of different data forms is needed [17].

Fig. 1 illustrates an architectural framework for big data analytics. Pooling data generated from multiple applications and locations (sources) is the first phase of big data analytics. In the second phase, the data is in a 'raw' state and needs to be cleansed and readied via several cleansing and transformation options, such as Extract, Transform, Load (ETL) steps [18]. Another approach, which works for the batch processing mode, is data warehousing, wherein data from diverse sources is cleansed, aggregated and made ready for processing [17]. Once the data is cleansed, it should replace the dirty data in the original sources to give legacy applications the improved data. In the third phase, and depending on whether the data is structured or unstructured, various data formats can be input to big data analytics platforms for processing, such as Hadoop [3] and MapReduce [19].

In this work, we introduce an MILP model to investigate the impact of the veracity of big data on network power consumption in bypass IP over WDM core networks. As a result, we make the following contributions: (i) we study the influence of *Veracity* by performing cleansing and backup for big data *Chunks* before processing, where a Backup Node (BN) location is optimally selected to store a copy of the cleansed big data *Chunks*; (ii) we perform our green edge, intermediate, and centralized processing technique to process the cleansed *Chunks* in optimal locations in the core network and compare the results to the classical approach that lacks progressive processing. The optimally selected processing locations for the green approach are either source PNs (SPNs) where *Chunks* are generated, inside location optimized DCs or at intermediate PNs (IPNs) between SPNs and DCs. Accordingly, the network elements, e.g., router ports, the routing paths, and the processing resources, are energy efficiently utilized to jointly minimize the power consumption of the overall network and processing resources.

The remainder of this paper is organized as follows: the rest of the current section presents the concept of greening big data networks and compares it with the classical big data networks using a clarifying example. In Section II, an MILP model for the veracity dimension is introduced and its results are presented. Section III concludes the paper.

Note that building PNs and BNs in our approach will add a capital cost compared to the classical approach. Capital expenditure (capex) and operational expenditure or (opex) study is, however, beyond the scope of this paper and we save this issue for future work.

The motivation of this work is that much generated data, currently, is not analyzed or addressed to extract insights at all [3]. Growth in the speed of generating data, as well as growth in the volume of data and in the variety of data sources is causing reduction in the percentage of processed data of organizations due to the lack of resources and poor analysis tools [3]. Thus, a large amount of the data that is to be processed is either neglected, deleted or delayed. Hence, there is unnecessary networking power consumption, extra wastage of storage and bandwidth because of transferring raw data, which leads to increasing the financial and environmental costs. Fig. 2 shows a decrease in the ratio of processed data to the overall huge volume of big data created continuously [20].

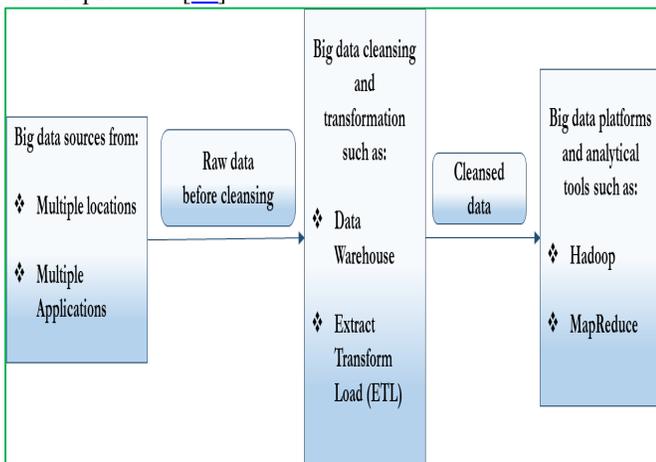

Fig. 1. Architectural framework for big data analytics [*17*].

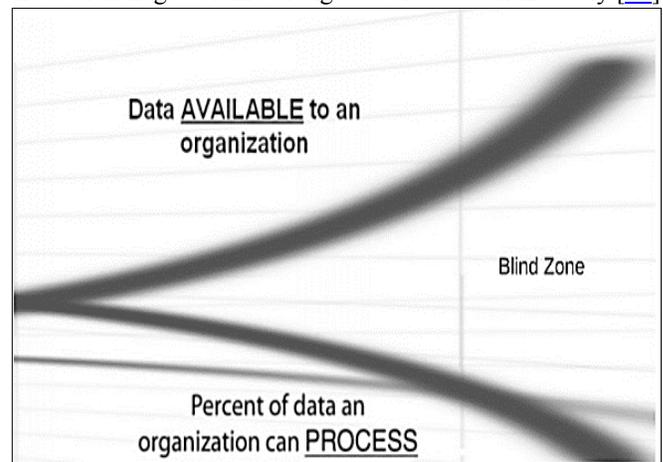

Fig 2. Volume of data is increasing, while percentage of data that can be processed is declining [20].

Managing massive data volumes calls for new processing and communications approaches. In addition, gathering and transmitting big data is exposing new challenges in terms of how to efficiently and economically transport big data over the network with acceptable service quality while providing adequate processing and storage resources. For instance, medical sensor data has to be transferred and processed in a very tight timeframe and the results have to be sent back to the hospital or patient wearable device as soon as possible before a health risk materializes. Such application level constraints impose even more challenges and hard trade-offs on the energy efficiency that can be attained from optimizing big data processing and networking [13].

*A. Classical Big Data Networks Vs. Green Big Data Networks*

The concept of green big data networks is illustrated in Fig. 3. Fig. 3-a displays a classical big data network where the processing of big data *Chunks* is achieved inside DCs after being generated and forwarded by the source nodes (an example of a source node in Fig. 3-a is National Health Service (NHS) node #14). In the green big data network, shown in Fig. 3-b, IP over WDM core nodes are attached to PNs (e.g., node #12) that can process *Chunks* and extract useful knowledge such as transportation and weather trends. We refer to the extracted knowledge in this paper as *Info*. These *Info* pieces are optimally transferred through energy efficient paths from the PNs to the DCs. The structure of a PN is similar to the cloud structure presented in [9]. It consists of a limited number of servers, storage (to store *Chunks*) and internal switches and routers. A PN is capable of edge processing the locally generated data and the data generated by other nodes and forwarding the results (*Infos*) to the DCs. The capacity of a PN is limited by the available space to build the PN inside the network center. Note in Fig. 3-b that the data generated by the source nodes can either be *Chunks*, *Infos* or both. The latter is the case if the source has processing elements. This type of source core node is referred to as a "Source PN" (SPN). On the other hand, the processing capability located at intermediate core nodes is referred to as "Intermediate PN" (IPN). IPNs perform the progressive processing for *Chunks* generated by other SPNs that did not perform local processing due to insufficient processing resources.

We assume, for realistic considerations, that each PN's processing and storage capacity varies from one PN to another. The DCs' capacities are, however, large enough for the central storing and processing of the remotely forwarded *Chunks* from the PNs. When *Chunks* are processed inside DCs, the corresponding results (i.e., *Infos*) are stored there. It is essential to mention here that the amount of computing resources required to process the *Chunks* in both approaches (i.e., in green big data networks and in classical networks) remains the same. In addition, in both approaches, the processing resources are utilized in an energy efficient manner where we calculate the processing power consumption assuming utilizing the minimum number of servers and also by employing slicing techniques [21]. However, the energy savings obtained from the green approach come from the optimal distribution of the processing resources among the network core nodes.

In big data analytics, because of the variety of big data applications, the ratio of the size of *Infos* to the size of *Chunks* (i.e., output/input) is diverse [11]. For example, *Infos* size is much smaller than *Chunks* size (i.e., the ratio can be << 1) in many big data applications, such as video monitoring in surveillance cameras to capture points of interest. While on the other hand, *Chunks* size approaches to *Infos* size (i.e., the ratio ≈ 1) would be in other big data job types, such as Call Detail Record (CDR) data produced by a telephone exchange or other telecommunications equipment. Further, there are several mixtures of jobs performed in big data analytics where the ratio is in between. In equation (1), we refer to this ratio as the Processing Reduction Ratio (PRR). Accordingly, the PRR is the ratio of the volume of *Infos* to the volume of *Chunks*. We assume that the volume of the *Chunks* is multiplied by different PRRs to produce the *Infos* carried by those *Chunks*. Therefore, big data traffic is significantly reduced in most big data analyses each time the data is processed progressively in the network before reaching the DCs. For instance, *Chunk* of 10 gigabits (Gb) and PRR of 0.001 results in *Info* of 10 megabits (Mb). Thus

$$Volume\ of\ Info = PRR \times Volume\ of\ Chunk. \quad (1)$$

To provide a clear picture of the relation between the output to the input size related to different big data jobs, we summarize the table that appeared in [11], which is obtained from a Facebook cluster of a MapReduce trace file within a two week period as shown in Table I.

TABLE I
MAPREDUCE FACEBOOK CLUSTER SUMMARY [*11*].

| Job counter | *Chunks* (input) size | *Infos* (output) size | PRR |
|---|---|---|---|
| 1145663 | 6.9 MB | 60 KB | 0.0086 |
| 11491 | 1.5 TB | 2.2 GB | 0.0014 |
| 670 | 2.1 TB | 2.7 GB | 0.0012 |
| 1876 | 711 GB | 860 GB | 1.21 |
| 169 | 2.7 TB | 260 GB | 0.096 |

Note that the network in Figure 2-b is greened by using a lower amount of resources when transmitting big data. For example, if the we transmit a chunk of 320 Gb (at a rate of 320 Gb/s), this needs 8 router ports (where each router port operates at 40 Gb/s) and consumes a total power of 6600 W (each router port consumes 825 W[22]), however, if this chunks carries an info of 40 Gb (and is transmitted at 40 Gb/s), then the consumed power is significantly reduce to 825 W, which is the power consumption of a single 40 Gb/s router port.

The Khazzoom-Brookes postulate argues that "energy efficiency improvements that, on the broadest considerations, are economically justified at the microlevel, lead to higher levels of energy consumption at the macrolevel." [23]. The human population is increasing and the number of Internet-connected devices are growing dramatically, which is expected to be 125 billion in 2030 [24]. Unlike the general domain of energy efficiency improvement, Information and Communications Technologies (ICT) have favourable features in relation to the Khazzoom-Brookes postulate – the "rebound effect". Improving the ICT energy efficiency may lead to higher uptake of ICT leading potentially to an increase in carbon emissions which we minimize through our work in this paper for example. More importantly however, ICT has the potential to reduce carbon emissions in other sectors such as the transport, manufacturing and agriculture sectors by introducing for example more efficient journeys, more efficient manufacturing processes and more efficient agriculture and watering systems enabled by ICT. Here higher uptake of ICT driven by improved ICT energy efficiency leads to higher carbon foot print savings in other sectors. In fact the SMARTer 2030 report shows that ICT can help reduce the global carbon foot print by an amounts equal to 10 times the carbon foot print of ICT by introducing efficiencies mainly in the transport, manufacturing and agriculture sectors [25].

Note that we did not consider big data job types where PRRs > 1 (i.e., *Chunk* size greater than *Info* size) since our main objective is to reduce the network power consumption and it is axiomatic that such a type of job will directly be forwarded to the DCs and skip our PNs. Forwarding such *Chunks* directly to the DCs will confine the extra traffic generated by the *Infos* to the inside of the DCs only.

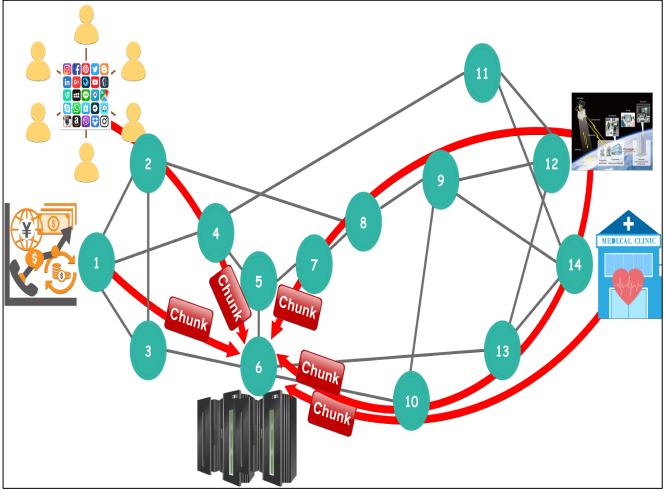

Fig. 3-a. Classical big data network [10].

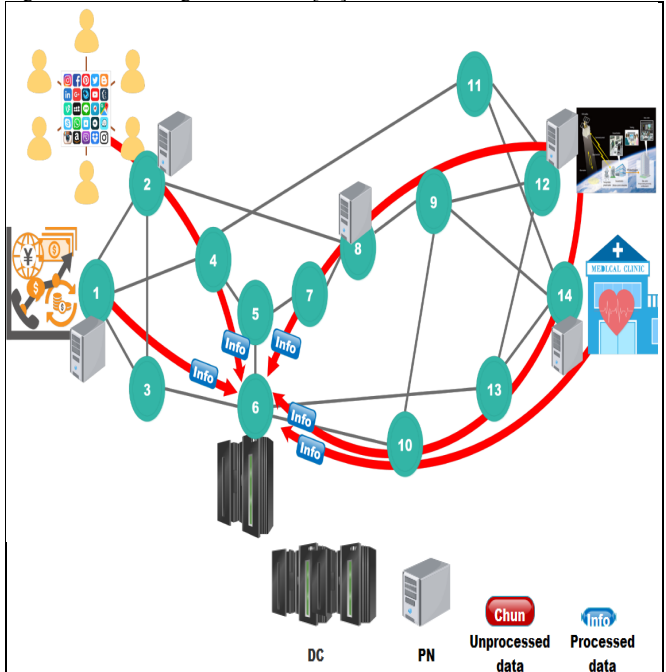

Fig. 3-b. Green big data network [10].

### B. Greening Big Data Networks: An Example

To demonstrate the concepts we propose in this work, consider the example network shown in Fig. 4. There are four zones in Fig. 4, with each connected to a certain PN, where each PN receives a different number of *Chunks* depending on its zone user population. For instance, zone 2 generates more *Chunks* compared to zone 4 that has a lower user population. The PN connected to a certain zone is referred to as a source PN (SPN) as it is the first PN in which *Chunks* are received from its corresponding zone and locally or centrally processed. Each SPN can locally process a different maximum number of *Chunks* depending on its processing, storage and internal switches and routers capacity. The remaining *Chunks* that cannot be processed locally in an SPN are forwarded either to another optimally selected PN or a DC. Those PNs that receive *Chunks* from other SPNs are called intermediate PNs (IPNs). An IPN, with respect to a given SPN, might itself be an SPN that implements local processing for its corresponding zone. This means that a PN can perform both the roles of SPN and IPN if needed. The unprocessed *Chunk* traffic from SPNs to IPNs or to DCs is called *Chunk* Big data Traffic (**CHT**). After processing the *Chunks* either in SPNs, IPNs or in the DCs, knowledge is extracted in the form of smaller rate traffic that we call the Info Big Data Traffic (**INF**). **INF** propagates from PNs (SPN or IPNs) toward DCs through the core network. Note that DCs have the special property that both the locally generated **INF** and the remotely received **INF** from other PNs do not flow outside them. As mentioned before, a PN is built at a certain core node; therefore, the PN ID is the same as the core node ID at which it is installed. This also applies to the DC ID. Each zone in Fig. 4 represents a probable scenario that our approach can optimize as follows:

**Zone 1**: The SPN #1 of zone 1 is capable of processing all incoming *Chunks* (*Chunks* #1, #2, #3) and all the output (*Infos* #1, #2, #3) are optimally aggregated to DC #1. This scenario generates only **INF** in the network from SPNs to DCs.

**Zone 2**: The SPN #2 of zone 2 can process *Chunks* #4, #5 and #6. *Chunk* #7 is, however, transported as a **CHT** to an optimal IPN (IPN #5) as one or more of the the resources (CPU, storage, internal switches, and routers) of SPN #2 have been fully utilized. After *Chunk* #7 is forwarded to IPN #5, it will be processed there and the output (*Info* #7) will be aggregated as an **INF** through an energy efficient route to DC #1.

**Zone 3**: The SPN #3 of zone 3 processes its own data (*Chunks* #8 and #9) and also acts as an IPN to process other incoming *Chunks* (*Chunk* #11 from SPN #4) when it is not being fully utilized. The movement of *Infos* from this PN represents the **INF**.

**Zone 4**: The SPN #4 of zone 4 has the smallest processing and storage space, thus it processes the smallest number of *Chunks* (*Chunk* #10) and forwards any extra *Chunks* to the next optimal PN or DC. For instance, *Chunk* #11 is forwarded to IPN #3. However, when all other PNs deplete their processing resources, then any extra unprocessed *Chunks* by SPN #4 (i.e., *Chunk* #12) will be uploaded directly from SPN #4 to be processed by an optimally selected DC (DC #2 in Fig. 4). For such an event, **CHT** starts to dominate the network traffic from SPNs to DCs.

## II. Veracity MILP Model

In this section, and for the completion of our work in [10, 12, 13] and [26], we introduce an MILP model for green big data networks by taking into consideration the bypass approach in an IP over WDM network. We placed capacitated PNs at each core node of the NSFNET, as depicted in Fig. 5, with DCs with large enough capacities. The NSFNET network consists of 14 nodes connected by 21 bidirectional links [27]. The DCs are used to process all extra big data *Chunks* originated by other PNs and to receive the results of the processed *Chunks* (i.e., *Infos*) produced by the PNs to store them for further use. The IP over WDM power consumption comprises the power consumption of the router ports, transponders, EDFAs, regenerators and optical switches.

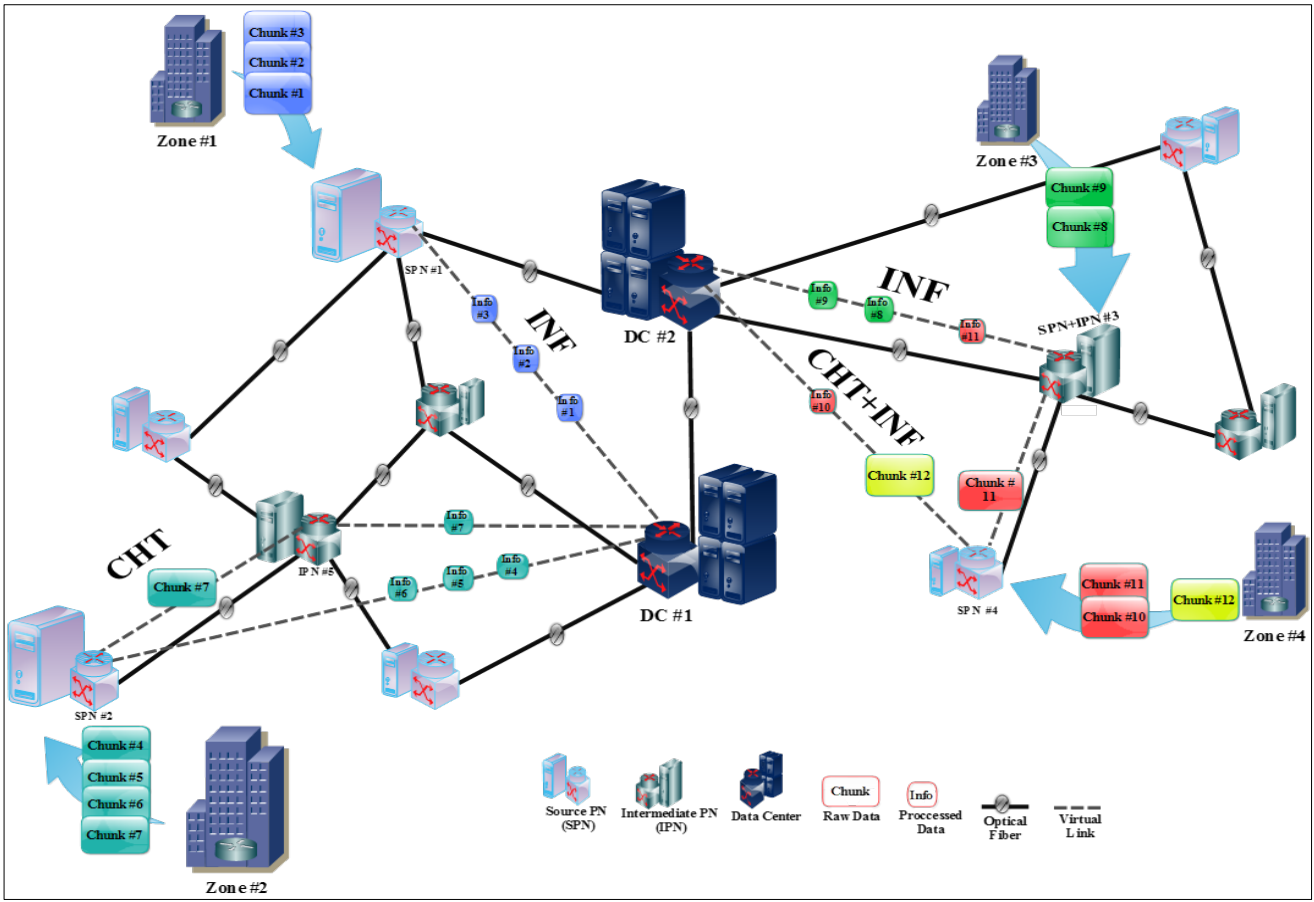

Fig. 4. Greening big data networks: an illustrative example.

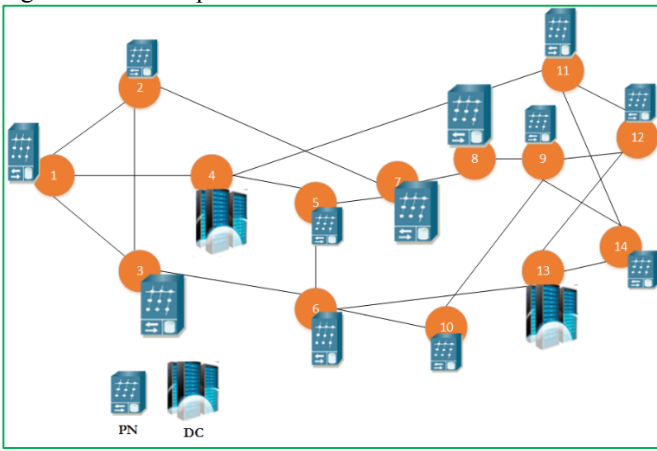

Fig. 5. NSFNET network with PNs.

Note that we assumed that there is additional traffic between core nodes, in addition to big data traffic, which is referred to as regular traffic [28].

The objectives of the present section are as follows: (i) to optimize the storage location of cleansed big data *Chunks* before processing, subject to PN storage limitations, and (ii) to build 1 to $N$ ($N$ is the number of nodes in the network) DCs for protecting the data by backing up one copy of each cleansed *Chunk* and to recall them for future analytics, and (iii) perform the our energy efficient edge, intermediate, and central processing technique while meeting the first two objectives. The optimization can find the location of 1 to $N$ such DCs. In the results, we evaluated the case of one backup DC for all the cleansed data, we refer to this DC as a Backup Node (BN).

We introduce the model to satisfy those objectives. The cleansing process is performed at each SPN when receiving raw data from multiple sources and generates cleansed *Chunks* with smaller volumes. These cleansed *Chunks* are progressively processed in the green big data network. Limited storage capacity for the PNs is considered to capture the distinct impact of storage limitations on the optimal location to store the cleansed *Chunks* in the network. We assume that the cleansing phase in the SPNs is implemented using temporary storage shared among the *Chunks*; however, the long term storage of the cleansed *Chunks* is determined by the model.

Note that the cleansing process is implemented offline at the source nodes to cleanse the actual data before being forwarded to the core node. A cleansed *Chunk* replica is a backup *Chunk* created in the event that the original *Chunk* is lost or destroyed. This backup *Chunk* is optimally stored in the BN. This BN could be either SPN or IPN. However, we do not consider co-locating the BN is one of the optimally placed DCs in this work. The number of employed BNs can be decided according to the level of resilience desired for the big data original *Chunks*.

**The following parameters are defined in the green big data networks model:**

- $SW_{sc}$    The CPU workload of the server required to process *Chunk c* generated at source node *s* (GHz).
- $PSB$    BN storage power per Gigabit (W/Gb).
- *s and d*    Denote source and destination points of regular traffic demand between a node pair.
- *m and n*    Denote end points of a physical fiber link in the optical layer.
- $R_{sd}$    The NSFNET regular traffic demand from node *s* to node *d* (Gbps).
- $N$    Set of IP over WDM nodes.
- $N_i$    The set of neighbor nodes of node *i* in the optical layer.
- $NS_p$    Number of servers at the PN *p*.
- $MSW$    Maximum server workload (GHz).
- $MP_p$    Maximum workload node *p*. $MP_p = NS_p*MSW$ (GHz).
- $MSR_p$    Maximum internal switch and router capacity of the PN *p* (Gbps).
- $MS_p$    Maximum storage of node *p* (Gb).
- $NCH$    Total number of *Chunks* in one node.
- $CH_s$    Set of *Chunks* in a source node *s*.
- $CHV_{sc}$    The volume of *Chunk c* generated at source node *s* (Gb).
- $PRR_{sc}$    Processing reduction ratio for *Chunk c* generated by node *s* (unitless).
- $WL$    Number of wavelengths in a fibre.
- $B$    Wavelength bit rate (Gbps).
- $S$    Maximum span distance between neighbouring EDFAs (km).
- $PR$    Power consumption of a router port (W).
- $PTR$    Power consumption of a transponder (W).
- $PO_i$    Power consumption of optical switch installed at node $i \in N$ (W).
- $PE$    Power consumption of EDFA (W).
- $PRG$    Power consumption of a regenerator (W).
- $D_{mn}$    Distance between node pair (m, n) (km).
- $A_{mn}$    Number of EDFAs on physical link (m, n). Typically, $A_{mn} = \left\lfloor \frac{D_{mn}}{S} - 1 \right\rfloor + 2$ [27].
- $RG_{mn}$    Number of regenerators on physical link (m, n).
- $PUN$    Power usage effectiveness of IP over WDM networks (unitless). PUN is defined as the ratio of the power drawn from the electric source to the power used by the equipment (networking in this case). PUN accounts for cooling, lighting and related power consumption.
- $PU$    Power usage effectiveness of the PNs and DCs (unitless).
- $SMP$    Server maximum power consumption (W).
- $SEB$    PNs' and DCs' switch energy per bit (W/Gbps).
- $REB$    PNs' and DCs' router energy per bit (W/Gbps).
- $RS$    Internal PNs' and DCs' switches redundancy.
- $RR$    Internal PNs' and DCs' routers redundancy.
- $RSG$    PNs and DCs storage redundancy.
- $PSG$    PNs' and DCs' storage power per Gigabit (W/Gb).
- $\delta$    Server power per GHz, $\delta = SMP / MSW$ (W/GHz). GHz is used to specify the capability of a processor and the number of processors a job needs.
- $DCN$    Number of location optimized DCs.

**The following variables are defined:**

- $CHT_{sp}$    Big data *Chunks* traffic generated at SPN *s* and directed to destination node *p* (*p* could be SPN, IPN or DC) (Gbps).
- $BN_d$    $BN_d = 1$ if node *d* is a backup node, else $BN_d = 0$.
- $BCH_{sd}$    Backup *Chunks* traffic from source node *s* to backup node *d*.
- $BCH_{ij}^{sd}$    Traffic flow of the backup *Chunk* traffic $BCH_{sd}$ between node pair (*s, d*) traversing virtual link (*i, j*).
- $AB_i$    Number of aggregation ports in router *i* utilized by backup *Chunk* traffic $BCH_{sd}$.
- $SBCH_d$    Amount of backup *Chunks* stored in BN *d* in Gb.
- $INF_{pd}$    Aggregated big data info traffic from PN *p* to DC *d*. Node *p* could be SPN or IPN only (Gbps).
- $C_{ij}$    Number of wavelength channels in the virtual link (*i,j*).
- $R_{ij}^{sd}$    Traffic flow of the regular traffic $R_{sd}$ between node pair (*s, d*) traversing virtual link (*i, j*).
- $W_{mn}^{ij}$    Number of wavelength channels in the virtual link (*i, j*) traversing physical link (*m, n*).
- $W_{mn}$    Number of wavelength channels in the physical link (*m,n*).
- $CHT_{ij}^{sp}$    Traffic flow of the big data *Chunks* traffic $CHT_{sp}$ between node pair (*s, p*) traversing virtual link (*i, j*).
- $INF_{ij}^{pd}$    Traffic flow of the big data info traffic $INF_{pd}$ between node pair (*p, d*) traversing virtual link (*i, j*).
- $AR_i$    Number of aggregation ports in router *i* utilized by regular traffic $R_{sd}$
- $ACH_i$    Number of aggregation ports in router *i* used in big data *Chunks* traffic $CHT_{sp}$.
- $AI_i$    Number of aggregation ports in router *i* utilized by big data *Info* traffic $INF_{pd}$.
- $F_{mn}$    Number of fibres in physical link (*m,n*).
- $PNW_p$    Total PN *p* workload (GHz).
- $Y_{spc}$    $Y_{spc} = 1$ if *Chunk c* is generated at SPN *s* and processed in PN *p*, else $Y_{spc} = 0$.
- $SCH_p$    Amount of big data *Chunks* stored in PN *p* (Gb).
- $DC_d$    $DC_d = 1$ if a DC is built at core node *d*, else $DC_d = 0$.

Under the bypass approach, the total IP over WDM network power consumption is composed of the following components

1) The power consumption of router ports

$$\sum_{i \in N} PR \cdot (AB_i + AR_i + ACH_i + AI_i) + PR \cdot \sum_{j \in N: i \neq j} (C_{ij}). \quad (2)$$

2) The power consumption of transponders
$$\sum_{m \in N} \sum_{n \in N_m} PTR \cdot W_{mn}. \quad (3)$$
3) The power consumption of regenerators is
$$\sum_{m \in N} \sum_{n \in N_m} PRG \cdot W_{mn} \cdot RG_{mn}. \quad (4)$$
4) The power consumption of EDFAs
$$\sum_{m \in N} \sum_{n \in N_m} PE \cdot A_{mn} \cdot F_{mn}. \quad (5)$$
5) The power consumption of optical switches
$$\sum_{i \in N} PO_i. \quad (6)$$

Equation (2) evaluates the total power consumption of the router ports for all the types of traffic, which are the regular traffic $R_{sd}$, big data *Chunks* traffic $CHT_{sp}$, big data info traffic $INF_{pd}$, and big data backup *Chunks* traffic $BCH_{sd}$. It computes the total power consumption of the ports aggregating data traffic and the ports connected to optical nodes. Equations (3) and (4) evaluate the power consumption of all the transponders and regenerators in the optical layer. Equation (5) evaluates the total power consumption of the EDFAs in the optical layer. Equation (6) evaluates the total power consumption of the optical switches. The power consumption of the PNs, DCs and the BN is composed of the following sections

6) The power consumption of internal PNs and DCs switches and routers

$$\begin{aligned} PSR = 2 \cdot \sum_{s \in N} \sum_{d \in N} BCH_{sd} \cdot (RS \cdot SEB + RR \cdot REB) \\ + \sum_{p \in N} \sum_{s \in N} CHT_{sp} \\ \cdot (RS \cdot SEB + RR \cdot REB) \\ + \sum_{p \in N} \sum_{d \in N} (CHT_{pd} + INT_{pd}) \\ \cdot (RS \cdot SEB + RR \cdot REB) \\ + \sum_{p \in N} \sum_{d \in N} INF_{pd} \\ \cdot (RS \cdot SEB + RR \cdot REB). \end{aligned} \quad (7)$$

Equation (7) calculates the power consumption of the internal switches and routers in the SPNs, IPNs and DCs, as well as the extra internal switches' and routers' power consumption in the SPNs and BNs resulting from sending backup *Chunks* between them. As we assume a homogeneous network and equipment, the total power consumption in the SPNs due to backup *Chunk* traffic is equal to the power consumption of the BN receiving that traffic, hence the factor of two in equation (7). We performed the analysis by considering a network architecture where $RS = RR = 1$.

7) The power consumption of servers
$$\sum_{p \in N} \delta \cdot PNW_p. \quad (8)$$
8) The power consumption of the PNs and DCs storage
$$\sum_{p \in N} SCH_p \cdot RSG \cdot PSG. \quad (9)$$
9) The power consumption the BN storage

$$\sum_{d \in N} SBCH_d \cdot RSG \cdot PSB. \quad (10)$$

Equation (9) represents the server power consumption. Although the server power consumption is a function of the idle power, maximum power and CPU utilization [29], we consider only $\delta = SMP/MSW$ to calculate its power consumption. This yields a close approximation (when a large number of servers is considered) even when there is idle power in each server. The difference is only in the last powered on server. Note that in the PN and DC servers, each server in our case is either fully utilized or is off. Equations (9) and (10) represent the storage power consumption of PN $p$ and BN $d$, respectively. We performed the analysis by considering a network architecture where $RSG = 1$.

To assess the impact of the veracity on the green big data networks, we integrate the PNs' storage limitations and cleansed *Chunks* backup dimension with the objective function that optimizes the variety. We chose the variety model as it encapsulates the volume analysis and considers a generic data input. The model is defined as follows

**Objective: Minimize**

$$\begin{aligned} PUN \cdot \Bigg( \sum_{i \in N} PR \cdot (AB_i + AR_i + ACH_i + AI_i) + PR \\ \cdot \sum_{j \in N : i \neq j} (C_{ij}) \\ + \sum_{m \in N} \sum_{n \in N_m} PTR \cdot W_{mn} \\ + \sum_{m \in N} \sum_{n \in N_m} PRG \cdot W_{mn} \cdot RG_{mn} \\ + \sum_{m \in N} \sum_{n \in N_m} PE \cdot A_{mn} \cdot F_{mn} \\ + \sum_{i \in N} EO_i \Bigg) \\ + PU \cdot \Bigg( \sum_{p \in N} \delta \cdot PNW_p + 2 \\ \cdot \sum_{s \in N} \sum_{d \in N} BCH_{sd} \cdot (RS \cdot SEB + RR \cdot R \\ + \sum_{p \in N} \sum_{s \in N} CHT_{sp} \\ \cdot (RS \cdot SEB + RR \cdot REB) \\ + \sum_{p \in N} \sum_{d \in N} (CHT_{pd} + INT_{pd}) \\ \cdot (RS \cdot SEB + RR \cdot REB) \\ + \sum_{p \in N} \sum_{d \in N} INF_{pd} \cdot (RS \cdot SEB + RR \cdot \\ + \sum_{p \in N} SCH_p \cdot RSG \cdot PSG \\ + \sum_{d \in N} SBCH_d \cdot RSG \cdot PSB \Bigg). \end{aligned} \quad (11)$$

Equation (11) presents the model objective, which is to minimize the IP over WDM network power consumption, the PN and DCs power consumption, and the BN power consumption.

**Subject to:**

**PNs, DCs, and BN Constraints:**

1) Processing counter of big data *Chunks* constraint

$$\sum_{p \in N} Y_{spc} = 1 \quad (12)$$
$$\forall s \in N, \forall c \in CH_s.$$

Constraint (12) ensures that a *Chunk* $c$ generated by PN $s$ is processed by no more than one PN $p$. However, our model performs slicing, i.e., multiple servers could process a given *Chunk* in a PN as long as these servers belong to that PN.

2) Big data *Chunks* traffic constraint

$$CHT_{sp} = \sum_{c \in CH_s} CHV_{sc} \cdot Y_{spc} \quad (13)$$
$$\forall s, p \in N.$$

Constraint (13) calculates the big data *Chunks* traffic generated at source node $s$ and directed to node $p$. This traffic is generated by transmitting $Chunk_{sc}$ from node s to node p in one second.

3) Aggregated processed big data traffic constraint

$$\sum_{d \in N} INF_{pd} = \sum_{s \in N} \sum_{c \in CH_s} CHV_{sc} \cdot Y_{spc} \cdot PRR_{sc} \quad (14)$$
$$\forall p \in N.$$

Constraint (14) represents the aggregated big data info traffic $INF_{pd}$ generated by PN $p$ and destined to DC $d$. The big data info traffic is obtained by multiplying the *Chunks* traffic incoming to the PN $p$ by the processing reduction ratio $PRR_{sc}$.

4) Number and locations of DCs constraints

$$\sum_{p \in N} INT_{pd} \geq DC_d \quad (15)$$
$$\forall d \in N,$$

$$\sum_{p \in N} INT_{pd} \leq Z \cdot DC_d \quad (16)$$
$$\forall d \in N, \text{ and}$$

$$DCN = \sum_{d \in N} DC_d. \quad (17)$$

Constraints (15) and (16) build a DC in location $d$ if that location is selected to store the results of the processed big data traffic (i.e., *Infos*) or selected to process the incoming big data *Chunks* from PNs, where $Z$ is a large enough unitless number to ensure that $DC_d = 1$ when $\sum_{p \in N} INF_{pd}$ is greater than zero. Constraint (17) limits the total number of built DCs to DCN.

5) PNs and DCs workload and processing capacity constraints

$$PNW_p = \sum_{s \in N} \sum_{c \in CH_s} SW_{sc} \cdot Y_{spc} \quad (18)$$
$$\forall p \in N \text{ and}$$
$$PNW_p \leq NS_p \cdot MSW + (M \cdot DC_p) \quad (19)$$
$$\forall p \in N.$$

Constraints (18) represents the total workload at PN $p$, which is the summation of the CPU workload of all the servers in that PN. Constraint (19) ensures that the total workload of PN $p$ will not exceed the maximum workload assigned to this PN, $M$ is a large enough unitless number. However, the workload capacity is large enough if a DC is built at core node $p$. Note that, the model implements a consolidation process by processing as many *Chunks* as possible within the same server to minimize the network power consumption and number of active servers.

6) PNs and DCs storage constraints

$$SCH_p = \sum_{s \in N} \sum_{c \in CH_s} CHV_{sc} \cdot Y_{spc} \quad (20)$$
$$\forall p \in N \text{ and}$$
$$SCH_p \leq MS_p + (H \cdot DC_p) \quad (21)$$
$$\forall p \in N.$$

Constraint (20) represents the size of *Chunks* in Gb stored in PN $p$. Constraint (21) ensures that the total data stored in PN $p$ does not exceed the storage capacity of that PN. $H$ is a large enough unitless number to guarantee that there is no storage capacity limitation at the DCs.

7) PNs and DCs internal switches and routers constraint

$$\sum_{s \in N} CHT_{sp} \leq MSR_p + (A \cdot DC_p) \quad (22)$$
$$\forall p \in N.$$

Constrain (22) ensures that the total amount of big data traffic between the PNs will not exceed the maximum switching and routing capacity of the internal switches and routers in those PNs. On the other hand, the capacity of the DCs' switches and routers is unlimited, where $A$ is a large enough unitless number to guarantee that there is no capacity limitation at the DCs. To avoid blocking of big data *Chunks*, we assume that the internal switches and routers capacity of the PNs is also large enough.

8) BN for big data *Chunks* constraint

$$\sum_{d \in N} BCH_{sd} = \sum_{s \in N} \sum_{c \in CH_s} CHV_{sc}. \quad (23)$$

Constraint (23) calculates the backup *Chunk* traffic generated at source node $s$ and stored at BN $d$. This is done by summing the individual cleansed original *Chunks* generated at source node $s$ that are to be stored at node $d$. This constraint ensures that only a single copy of a *Chunk* will be stored.

9) Number of BNs and location constraints

$$\sum_{s \in N} BCH_{sd} \geq BN_d \quad (24)$$
$$\forall d \in N,$$

$$\sum_{s \in N} BCH_{sd} \leq Z \cdot BN_d \quad (25)$$
$$\forall d \in N,$$

$$BN = \sum_{d \in N} BN_d = 1 \text{ and} \quad (26)$$

$$DC_d \leq 1 - BN_d \quad (27)$$
$$\forall d \in N.$$

Constraints (24) and (25) build a BN in location $d$ if that location is selected to store the backup *Chunks*, where $Z$ is a large enough unitless number to ensure that $BN_d = 1$ when $\sum_{s \in N} BCH_{sd}$ is greater than zero. Constraint (26) calculates the total number of backup nodes in the network (we display the results for when only one backup node is to be optimally selected in the network). Constraint (27) ensures that selecting a node as a DC and BN is not allowed.

10) BN and PNs storage capacity constraint

$$SBCH_d \leq MS_d + H \cdot BN_d \quad (28)$$
$$\forall d \in N.$$

Constraint (28) ensures that if a PN $d$ is chosen to be a BN, then that node will have enough large storage capacity, where $H$ is a large enough unitless number, while the PN $d$ will have limited storage otherwise.

1) Amount of stored backup *Chunks* constraint

$$SBCH_d = \sum_{s \in N} \sum_{c \in CH_s} CHV_{sc} \cdot BN_d \quad (29)$$
$$\forall d \in N.$$

Constraint (29) represents the size of the backup *Chunks* stored in the BN $d$.

**The IP over WDM Network Constraints**

1) Flow conservation constraints for the regular traffic

$$\sum_{j \in N: i \neq j} R_{ij}^{sd} - \sum_{j \in N: i \neq j} R_{ji}^{sd} = \begin{cases} R_{sd} & i = s \\ -R_{sd} & i = d \\ 0 & otherwise \end{cases} \quad (30)$$
$$\forall s, d, i \in N: s \neq d.$$

2) Flow conservation constraints for the big data *Chunks* traffic

$$\sum_{j \in N: i \neq j} CHT_{ij}^{sp} - \sum_{j \in N: i \neq j} CHT_{ji}^{sp} = \begin{cases} CHT_{sp} & i = s \\ -CHT_{sp} & i = p \\ 0 & otherwise \end{cases} \quad (31)$$
$$\forall s, p, i \in N: s \neq p.$$

3) Flow conservation constraints for the big data *Info* traffic

$$\sum_{j \in N: i \neq j} INF_{ij}^{pd} - \sum_{j \in N: i \neq j} INF_{ji}^{pd} = \begin{cases} INF_{pd} & i = p \\ -INF_{pd} & i = d \\ 0 & otherwise \end{cases} \quad (32)$$
$$\forall p, i \in N, \forall d \in N: p \neq d.$$

Constraints (30-32) represent the flow conservation constraints for the regular traffic $R_{sd}$, big data *Chunks* traffic $CHT_{sp}$ and big data info traffic $INF_{pd}$, in the IP layer. These constraints ensure that the total outgoing traffic should be equal to the total incoming traffic, except for the source and destination nodes. It can also ensure that the flow can be divided into multiple flow paths in the IP layer.

4) Flow conservation constraints for big data backup *Chunks* traffic

$$\sum_{j \in N: i \neq j} BCH_{ij}^{sd} - \sum_{j \in N: i \neq j} BCH_{ij}^{sd} = \begin{cases} BCH_{sd} & i = s \\ -BCH_{sd} & i = d \\ 0 & otherwise \end{cases} \quad (33)$$
$$\forall s, i \in N, \forall d \in N: s \neq d.$$

Constraint (33) represents the flow conservation constraint for big data backup traffic in the IP layer. This constraint ensures the total outgoing traffic should be equal to the total incoming traffic, except for the source and destination nodes. It can also ensure that the flow can be divided into multiple flow paths in the IP layer.

5) Virtual link capacity constraint

$$\left( \sum_{s \in N} \sum_{d \in N: s \neq d} R_{ij}^{sd} + \sum_{s \in N} \sum_{d \in N: s \neq d} BCH_{ij}^{sd} \right.$$
$$+ \sum_{s \in N} \sum_{p \in N: s \neq p} CHT_{ij}^{sp}$$
$$\left. + \sum_{p \in N} \sum_{d \in N: p \neq d} INF_{ij}^{pd} \right) \leq C_{ij} \cdot B \quad (34)$$
$$\forall i, j \in N: i \neq j.$$

Constraint (34) ensures that the summation of all the traffic types flow through a virtual link and does not exceed its capacity.

6) Optical layer flow conservation constraints:

$$\sum_{n \in N_m} W_{mn}^{ij} - \sum_{n \in N_m} W_{mn}^{ij} = \begin{cases} C_{ij} & m = i \\ -C_{ij} & m = j \\ 0 & otherwise \end{cases} \quad (35)$$
$$\forall i, j, m \in N: i \neq j.$$

Constrain (35) represents the flow conservation constraints in the optical layer. It assumes that the total outgoing wavelengths in a virtual link should be equal to the total incoming wavelengths, except for the source and the destination nodes of the virtual link.

7) Physical link capacity constraints

$$\sum_{i \in N} \sum_{j \in N: i \neq j} W_{mn}^{ij} \leq WL \cdot F_{mn}. \quad (36)$$
$$\forall m \in N, n \in N_m.$$

Constraint (36) ensures that the summation of the wavelengths in a virtual link traversing a physical link do not exceed the capacity of the fibre in the physical link.

8) Wavelengths capacity constraint

$$\sum_{i \in N} \sum_{j \in N: i \neq j} W_{mn}^{ij} = W_{mn} \quad (37)$$
$$\forall m \in N, n \in N_m.$$

Constraint (37) ensures that the summation of the wavelengths traversing a physical link do not exceed the total number of wavelengths in that link.

9) Number of aggregation ports utilized by regular traffic constraint

$$AR_i = \frac{1}{B} \cdot \sum_{d \in N: i \neq d} R_{id} \quad (38)$$
$$\forall i \in N.$$

10) Number of aggregation ports utilized by **CHT** traffic constraint

$$ACH_i = \frac{1}{B} \cdot \sum_{p \in N: i \neq p} CHT_{ip} \quad (39)$$
$$\forall i \in N.$$

11) Number of aggregation ports utilized by **INF** traffic constraint

$$AI_i = \frac{1}{B} \cdot \sum_{d \in N: i \neq p} INF_{id} \quad (40)$$
$$\forall i \in N.$$

12) Number of aggregation ports utilized by **BCH** traffic constraint

$$AB_i = \frac{1}{B} \cdot \sum_{d \in N: i \neq d} BCH_{id} \quad (41)$$
$$\forall i \in N.$$

Constraints (38-41) calculate the number of aggregation ports for each router that serves the regular traffic $R_{sd}$, big data *Chunks* traffic $CHT_{sp}$ and big data info traffic $INF_{pd}$

### A. Results of Veracity Scenarios

Our MILP model was evaluated using the NSFNET network depicted in Fig. 5. The storage capacity of the PNs were assigned to be large enough between 10 Pb - 70 Pb. The number of servers per PN varied between 10 and 60 for all evaluated scenarios. Note that we used processor cycles in GHz as a measure of the total processing capability of a node [30]. Table II summarizes the input parameters to the model.

TABLE II
INPUT DATA FOR VOLUME MODEL.

| Parameter | Value |
| --- | --- |
| $PRR_{sc}$ | 0.01-1 (random uniform) |
| PNs storage capacity ($MS_p$) $\forall p \in N$ | 10 Pb - 70 Pb |
| Uncleansed *Chunk* volume in Gb ($Chunk_{sc}$) | 50-300 (random uniform) |
| Cleansed *Chunk* volume in Gb ($Chunk_{sc}$) | 10-220 (random uniform) |
| Number of *Chunks* per PN ($\beta$) | 10-60 |
| Number of servers per PN ($NS_p$) | 10-30 |
| CPU workload per *Chunk* in GHz ($W_{sc}$) | 1-4 (random uniform) |
| Server CPU capacity *in GHz* ($MSW$) | 4 GHz |
| Max server power consumption ($MSP$) | 300 W [9] |
| Energy per bit of the PNs and DCs switch ($SEB$) | 11.875 W/Gbps [9] |
| Energy per bit of the PNs and DCs router ($REB$) | 7.727 W/Gbps [9] |
| Storage power consumption ($PSG$) | 0.008 W/Gb [9] |
| IP over WDM router power consumption ($PR$) | 825 W [22] |
| IP over WDM regenerator power consumption ($PRG$) | 334 W [22] |
| IP over WDM transponder power consumption ($PTR$) | 167 W [22] |
| IP over WDM optical switch power consumption ($PO_i$) $\forall i \in N$ | 85 W [22] |
| IP over WDM EFDA power consumption ($PE$) | 55 W [22] |
| Wavelength bit rate ($B$) | 40 Gbps |
| Span distance between EDFAs ($S$) | 80 km |
| Number of wavelengths per fibre ($WL$) | 32 |
| Number of location optimized DCs ($DCN$) | 2 |
| IP over WDM power usage effectiveness ($PUN$) | 1.5 [9] |
| PNs and DCs power usage effectiveness ($PU$) | 2.5 [9] |

The MILP in this section is used to evaluate the proposed big data networks. In addition, the same model can be used to evaluate the classical approach by introducing a constraint that prevents the processing of big data outside the DCs. The classical model is characterized by:

- Each node of the NSFNET generates a similar number of *Chunks* as in the scenarios of the green big data networks model.
- The classical model optimally selects DC locations to host, store and process *Chunks* that are forwarded by all the nodes through energy efficient routes. Further, it selects an optimal BN location for storing the *Chunks*.
- The extracted knowledge from the *Chunks* (i.e. *Infos*) is stored in the same DC that processed the corresponding *Chunks* for further analysis.
- The DCs process the incoming *Chunks* with minimum power consumption by utilizing the lowest number of servers that can handle those chunks. This is emphasized by calculating the processing power consumption required by the minimum number of servers.
- The processing of big data is achieved inside the DCs only when there are no PNs in the network. The objective of the model is to minimize the network power consumption and the DC power consumption.

Note that a number of computational resources required to process the data is the same in our approach and the classical approach where all *Chunks* are processed inside DCs.

To provide a holistic assessment of the impact of the veracity dimension on green big data networks, we evaluate the proposed work in two scenarios as follows:

*A.1 Veracity with Large Enough Storage Capacity*

We compare the green big data approach to the classical approach where *Chunks* are not cleansed and directly sent to DCs. This means that the raw traffic volumes in the green big data network are smaller compared to the classical approach where the cleansing and processing happen inside the DCs only. For each approach, we evaluate two modes of operation. In the first mode there is a BN in the network and in the second mode, no BN is employed. Therefore, we can compare the two approaches (green vs. classical) against each other for each mode. For the green big data approach, the cleansed *Chunk* volumes vary in a random uniform distribution between 10 Gb and 220 Gb. On the other hand, a larger volume range for the classical approach is assumed between 50 Gb and 300 Gb. Note that we chose these volumes as they are close to the realistic values provided in [11]. In all cases, the SPNs storage is large enough to store the cleansed data. We used the input values shown in Table II to examine the influence of veracity on network power consumption.

Fig. 6-a shows the network power consumption of the classical approach and green big data networks approach with and without performing the *Chunks* backup modes. The system performance yields noteworthy differences in the network power saving between the two modes. For instance, the maximum power saving is 45% in the backup mode while it is 58% for the no backup mode at $\beta = 50$. The average power saving in the backup mode is 41% and 52% in the no-backup mode. The reason for the lower power savings in the backup mode is due to the presence of the extra backup traffic between the SPNs and BNs that increases the network power consumption and reduces the network power savings. On the other hand, there is no backup traffic in the no-backup scenario, but only **CHT** appears in the network, which is either from SPNs to IPNs or from SPNs to DCs, thereby minimizing the power consumption.

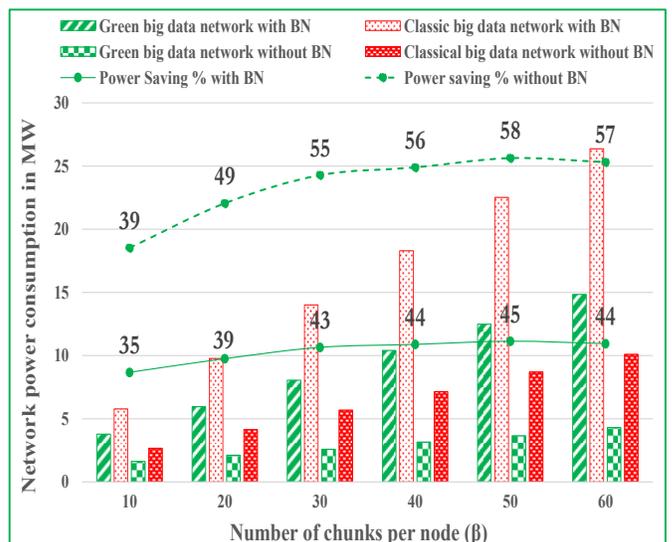

Fig. 6-a. Network power consumption for classical and green big data networks with and without BN for veracity scenario A.1.

Fig. 6-b displays the PN, DC and BN storage utilizations with different values of β for the green big data approach. It shows that node 6 is selected as the BN at all values of β. This is due to the strategic location of node 6, which has the minimum number of hops to all other nodes. In addition, the DC locations are selected at nodes 4 and 13 for all values of β. Note that up to β = 30, the DC storage utilization remains steady as the original *Chunks* are processed either locally in the SPNs or intermediately in the IPNs. At β = 40 the **BCH** dominates the network compared to the **CHT** and INF. At the stage where 40 < β ≤ 50, the DCs start to receive a considerable number of original *Chunks* because most PN resources are utilized. Accordingly, **CHT** increases considerably in addition to the existing **BCH**, thereby yielding an overall increase in network power consumption as discussed earlier.

When 50 < β ≤ 60, all PN processing resources are depleted, thus, the increase in the DC storage utilization is significant as any extra *Chunks* are forwarded to a DC for storing and processing. Consequently, the combined traffic (**CHT** and **BCH**) is now at its maximum value at β = 60.

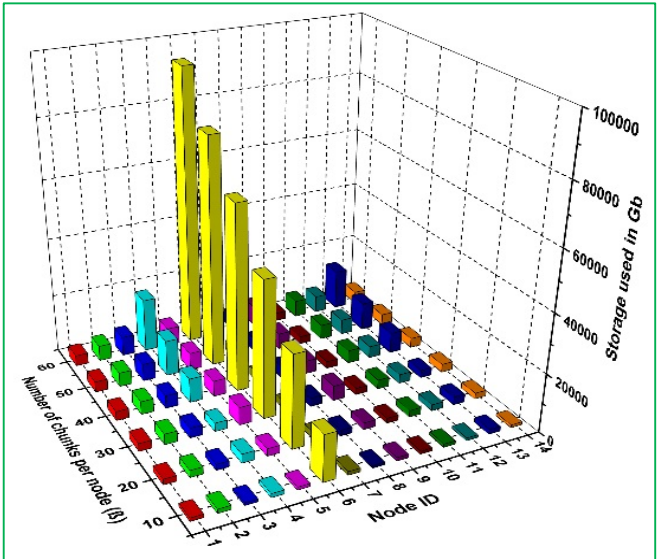

Fig. 6-b. Storage used in the PNs and DCs and BN with different values of β for veracity scenario A.1.

Note that since our model selected the BN location at node #6 under different big data traffic patterns, we suggest that this location can be fixed at that location as the best location, for building the BN, which introduces favourable outcomes in terms of minimizing both cost and power consumption. Furthermore, in the distributed energy efficient clouds over the core networks in [9], node #6 was optimally selected as the best location for the datacentre and storage under various analysing scenarios.

*A.1.1 Employing Renewable Energy Source*
We consider in this section on reducing the $CO_2$ emissions of Backbone IP over WDM networks. A MILP optimization model for hybrid-power" (i.e. renewable and non-renewable energy sources) IP over WDM networks was set up to minimize the non-renewable energy consumption and $CO_2$ emission. Typically a one square meter silicon solar cell can produce about 0.28 kW of power [31]. We assume that the maximum Solar energy available to a node is between 20 kW and 80 kW, therefore a total solar cell area of about 100 m² to 400 m² is required. The premises housing a core or access network node (Telecommunications office) is able typical to provide such surface area.

In addition to main parameters defined in Section II of the MILP model, the following parameter is used in the MILP model:

$SOL_s$    Maximum amount of solar power available to node *s* (kW).

In addition to the main variables defined in Section II of the MILP model, the following variables are introduced in this section:

$RE_s$    The amount of renewable power consumed by IP over WDM node *s* (kW).
$NRE_s$    The amount of non-renewable power consumed by IP over WDM node *s* (kW).
$NPC_s$    The total amount of renewable and non-renewable power consumed by IP over WDM node *s* (kW).
$TNRE$    The total amount of non-renewable power consumed by all the IP over WDM nodes in the network (kW).

In addition to the main constraints defined in Section II of the MILP model, the following constraints are introduced in this section:

$$NPC_s = RE_s + NRE_s \quad (42)$$

$$\forall s \in N.$$

$$RE_s \leq SOL_s \quad (43)$$

$$\forall s \in N.$$

Constraint (42) evaluates the total amount of renewable and non-renewable power consumed by node *s*. Constraint (43) ensures that the amount of the solar power consumed by node *s* does not exceed the maximum solar power available to that node.

$$TNRE = \sum_{s \in N} NRE_s \quad (44)$$

Constraint (44) determines the total amount of non-renewable power consumed by all the IP over WDM network equipment.

The objective function of the MILP model is updated to minimize the total non-renewable energy consumption of the network.

We evaluated the MILP in several cases at the point where 50 chunks are provided to each PN in the previous scenario (see Fig. 5a, the case of green big data network with BN at β=50). Fig. 7 shows that providing more solar power results in decreasing the non-renewable power consumption. For example, at the 20kW solar power case, the non-renewable power consumption is decreased by around 3% compared to the case where there is no solar power provided to the network. The network becomes greener by around 9% when providing 80kW of solar power to each node.

To conclude, although our approach achieved significant non-renewable power saving, employing renewable energy sources reduces the carbon footprint of the network by around 9% more.

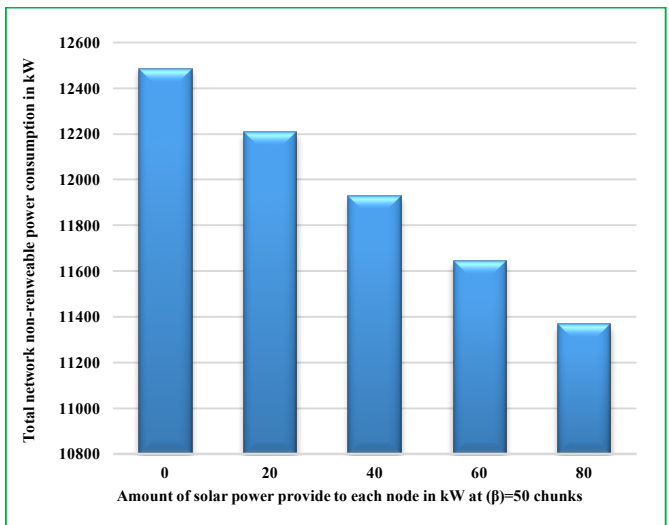

Fig. 7 Total non-renewable power consumption vs amount of solar power provided to the IP over WDM nodes at (β)=50 chunks.

*A.2 Veracity with Limited Storage Capacity per PN*
In this scenario, we consider the impact of using limited storage in the PNs to allow the model to optimize the location of the cleansed *Chunks* for the green big data approach. We reused the same inputs that appeared in Table II except for the PNs storage capacity as it is now limited between 1 Tb to 4 Tb per PN following a uniform distribution.

Fig. 8-a & b display the PN storage and processing utilization, respectively. To illustrate the impact of limited storage capacity on the green big data networks with cleansing, we show the results for two SPNs, #3 and #12. We assigned a low storage capacity of 1Tb to these two SPNs and a high processing capacity of 30 servers. Fig. 8-a shows that for all values of β, each of the two SPNs could store a maximum cleansed *Chunk* volume ≤ 1 Tb only, and any cleansed *Chunks* above 1 Tb would be optimally forwarded and stored at another IPN or to the DCs for processing. Recall that the SPNs have no cleansing limitations as the cleansing temporary storage is shared among raw *Chunks*. However, the cleansed *Chunks* need to be stored for long term usage at certain PNs, as optimally selected by the model. For example, at β = 60, the total cleansed volume was 7617 Gb at PN #6 and 6087 Gb at PN #12, however, the actual stored amount of data was 995 Gb inside PN #6 and 998 Gb inside PN #12, due to the 1 Tb storage capacity of both PNs. The remaining cleansed *Chunks* by those two SPNs were optimally sent and stored at one of the DCs.

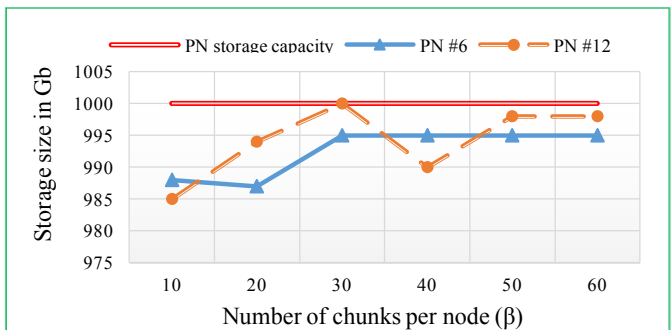

Fig. 8-a PNs storage size with different values of β for veracity scenario A.2.

Fig. 8-b shows the processing utilization for the two example PNs. Interestingly, for all values of β, the average processing utilizations of both PN #6 and PN #12 were around 16 GHz below the maximum processing capacity (which is 120 GHz: 30 servers with 4 GHz CPU per server). This is because the model skips those PNs after full utilization of the storage capacity regardless of the availability of processing resources, which leads to a smaller number of locally processed *Chunks*.

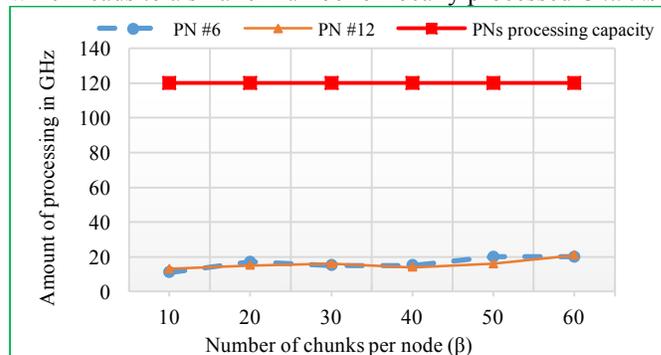

Fig. 8-b Utilization of processing capacity for different values of β when considering limited storage per PN for veracity scenario A.2.

Fig. 8-c illustrates the impact of considering limited PNs storage on network power consumption for the classical approach and the green big data networks approach with and without the cleansing of backup *Chunks*. The Figure shows a decrease in the network power saving for both backup and no-backup modes compared to the veracity scenario A.1 where the PNs have a large enough storage capacity. The maximum power saving obtained in this scenario declined to 40% for the backup mode and 51% for the no-backup mode at β = 50, while it was 45% and 58% at β = 50 with and without backup mode in the veracity scenario A.1, respectively. The reason behind this decrease in power saving is that the limited storage capacity of the PNs leads to a smaller number of cleansed *Chunks* being processed locally in the edge and progressively in the INPs regardless of the fact that there are still available processing resources. Thus, increasing the amount of **CHT** in the green big data networks results in a higher network power consumption. The average network power saving obtained in the present scenario for the backup mode was 38%, while it was 47% for the no-backup mode.

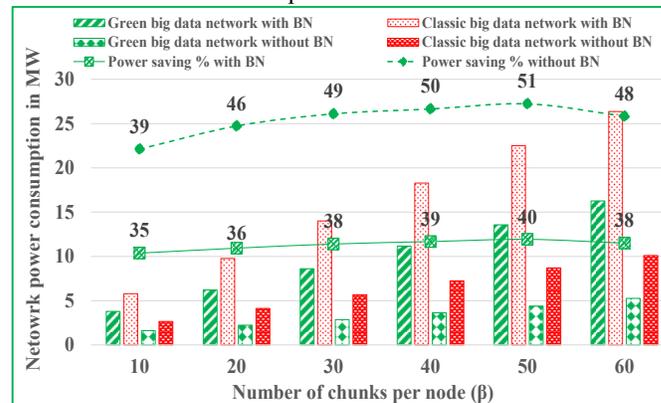

Fig. 8-c Network power consumption for classical and green big data networks with and without BN with limited storage per PN for veracity scenario A.2.

## III. CONCLUSIONS

This paper introduced a Mixed Integer Linear Programming (MILP) model to investigate the impact of the veracity of big data on greening big data networks in bypass IP over WDM core networks. We presented the green big data approach by introducing Processing Nodes (PNs) that are attached to the ISP network centers which host the IP over WDM nodes. A PN is a small version of a datacenter (DC) with a capacity that is limited by the available space to build the PN inside the network center. We introduced a progressive processing technique to serve big data applications in source PNs, intermediate PNs and DCs taking into consideration the veracity dimension. We optimized the storage locations of the cleansed data as well as optimizing the location of a single backup node to store a copy of the cleansed *Chunks* for future use. The veracity scenarios had an average network power savings of up to 52% in the no backup mode and up to 41% in the backup mode. The lower saving for the no backup mode is due to the movement of *Chunks* from the source PNs to the backup PN without processing them during that journey. In addition, we noted that the veracity scenario under the PNs storage limitations utilizes fewer of the available processing resources as it is influenced by the PNs limited storage capacity.

ACKNOWLEDGEMENT

The authors would like to acknowledge funding from the Engineering and Physical Sciences Research Council (EPSRC), INTERNET (EP/H040536/1) and STAR (EP/K016873/1). The first author would like to acknowledge the HCED Iraq and the University of Kerbala-Iraq for funding his scholarship. All data are provided in full in the results section of this paper.

References

[1] F. Brian, "Software crisis 2.0," *Computer,* vol. 45, pp. 89-91, 2012.
[2] J. Gantz and D. Reinsel, "The digital universe in 2020: Big data, bigger digital shadows, and biggest growth in the far east," *IDC iView: IDC Analyze the future,* vol. 2007, pp. 1-16, 2012.
[3] P. Zikopoulos and C. Eaton, *Understanding big data: Analytics for enterprise class hadoop and streaming data*: McGraw-Hill Osborne Media, 2011.
[4] D. Zeng, L. Gu, and S. Guo, "Cost minimization for big data processing in geo-distributed data centers," in *Cloud Networking for Big Data*, ed: Springer, 2015, pp. 59-78.
[5] I. Satoh, "MapReduce-Based Data Processing on IoT," in *Internet of Things (iThings), 2014 IEEE International Conference on, and Green Computing and Communications (GreenCom), IEEE and Cyber, Physical and Social Computing(CPSCom), IEEE*, 2014, pp. 161-168.
[6] C. Jayalath, J. Stephen, and P. Eugster, "From the cloud to the atmosphere: running mapreduce across data centers," *Computers, IEEE Transactions on,* vol. 63, pp. 74-87, 2014.
[7] Q. Xia, W. Liang, and Z. Xu, "Data locality-aware query evaluation for big data analytics in distributed clouds," in *Advanced Cloud and Big Data (CBD), 2014 Second International Conference on*, 2014, pp. 1-8.
[8] L. Rupprecht, "Exploiting in-network processing for big data management," in *Proceedings of the 2013 SIGMOD/PODS Ph. D. symposium*, 2013, pp. 1-6.
[9] A. Q. Lawey, T. E. El-Gorashi, and J. M. Elmirghani, "Distributed energy efficient clouds over core networks," *Journal of Lightwave Technology,* vol. 32, pp. 1261-1281, 2014.
[10] A. M. Al-Salim, A. Q. Lawey, T. El-Gorashi, and J. M. Elmirghani, "Energy Efficient Tapered Data Networks for Big Data Processing in IP/WDM Networks," in *Transparent Optical Networks (ICTON), 2015 17th International Conference on*, 2015, pp. 1-5.
[11] Y. Chen, A. Ganapathi, R. Griffith, and R. Katz, "The case for evaluating mapreduce performance using workload suites," in *Modeling, Analysis & Simulation of Computer and Telecommunication Systems (MASCOTS), 2011 IEEE 19th International Symposium on*, 2011, pp. 390-399.
[12] A. M. Al-Salim, H. M. Mohammad Ali, A. Q. Lawey, T. El-Gorashi, and J. M. Elmirghani, "Greening Big Data Networks: Volume Impact," in *Transparent Optical Networks (ICTON), 2016 17th International Conference on*, 2016, pp. 1-6.
[13] A. M. Al-Salim, A. Q. Lawey, T. E. El-Gorashi, and J. M. Elmirghani, "Energy efficient big data networks: impact of volume and variety," *IEEE Transactions on Network and Service Management,* 2017.
[14] A. Al-Salim, T. EL-Gorashi, A. Lawey, and J. M. Elmirghani, "Greening Big Data Networks: Velocity Impact," *IET Optoelectronics,* 2017.
[15] B. d. v. K. Normandeau, variety, velocity and veracity", insideBIGDATA, 2016. [Online]. Available: http://insidebigdata.com/2013/09/12/beyond-volume-variety-velocity-issue-big-data-veracity/. [Accessed: 09- Nov- 2016].
[16] E. Rahm and H. H. Do, "Data cleaning: Problems and current approaches," *IEEE Data Eng. Bull.,* vol. 23, pp. 3-13, 2000.
[17] W. Raghupathi and V. Raghupathi, "Big data analytics in healthcare: promise and potential," *Health Information Science and Systems,* vol. 2, pp. 1-10, 2014.
[18] B. Larson, *Delivering Business Intelligence*: New York, 2009.
[19] J. Dean and S. Ghemawat, "MapReduce: simplified data processing on large clusters," *Communications of the ACM,* vol. 51, pp. 107-113, 2008.
[20] P. C. Zikopoulos, C. Eaton, D. DeRoos, T. Deutsch, and G. Lapis, *Understanding Big Data – Analytics for Enterprise Class Hadoop and Streaming Data*. McGraw-Hill: Aspen Institute, 2012.
[21] H. Goudarzi and M. Pedram, "Energy-efficient virtual machine replication and placement in a cloud computing system," in *Cloud Computing (CLOUD), 2012 IEEE 5th International Conference on*, 2012, pp. 750-757.
[22] GreenTouch, "GreenTouch Final Results from Green Meter Research Study Reducing the Net Energy Consumption in Communications Networks by up to 98% by 2020," *A GreenTouch White Paper,* vol. Version 1, 15 August 2015.
[23] M. Einhorn, "Economic implications of mandated efficiency standards for household appliances: an extension," *The Energy Journal,* vol. 3, pp. 103-109, 1982.
[24] IHSMarket, "The Internet of Things is here and growing exponentially".[Online]. Available: https://technology.ihs.com/596542/number-of-connected-iot-devices-will-surge-to-125-billion-by-2030-ihs-markit-says. [Accessed: 29-May-2018]."
[25] A. Strategy, "# SMARTer2030: ICT solutions for 21st century challenges," *The Global eSustainability Initiative (GeSI), Brussels, Brussels-Capital Region, Belgium, Tech. Rep,* 2015.
[26] A. M. Al-Salim, T. El-Gorashi, A. Q. Lawey, and J. M. Elmirghani, "Greening Big Data Networks: Velocity Impact," *Journal of Transactions on Big Data,* Submitted on Nov. 2016.
[27] G. Shen and R. S. Tucker, "Energy-minimized design for IP over WDM networks," *Journal of Optical Communications and Networking,* vol. 1, pp. 176-186, 2009.
[28] X. Dong, T. E. El-Gorashi, and J. M. Elmirghani, "On the energy efficiency of physical topology design for IP over WDM networks," *Journal of Lightwave Technology,* vol. 30, pp. 1694-1705, 2012.
[29] A. Beloglazov, J. Abawajy, and R. Buyya, "Energy-aware resource allocation heuristics for efficient management of data centers for

cloud computing," *Future generation computer systems,* vol. 28, pp. 755-768, 2012.

[30] N. B. Rizvandi, J. Taheri, R. Moraveji, and A. Y. Zomaya, "On modelling and prediction of total CPU usage for applications in mapreduce environments," in *Algorithms and Architectures for Parallel Processing*, ed: Springer, 2012, pp. 414-427.

[31] A. Wang, J. Zhao, and M. Green, "24% efficient silicon solar cells," *Applied physics letters,* vol. 57, pp. 602-604, 1990.